\documentclass[twocolumn,pra,aps,showpacs]{revtex4}

\usepackage{mathptmx}
\usepackage{amsthm} 
\usepackage{subfigure}
\usepackage{psfrag,graphicx}
\usepackage{dcolumn}
\usepackage{amsmath,amssymb}
\usepackage{bm}
\usepackage[dvips]{color}
\usepackage{overpic}

\usepackage[colorlinks=true,citecolor=myblue,linkcolor=mygrey]{hyperref}
\definecolor{myblue}{named}{MidnightBlue}

\definecolor{mygrey}{rgb}{0.06,0.31,0.55} 
\definecolor{mygreen}{rgb}{0.85,1,0.9}
\definecolor{myzard}{cmyk}{0,0,0.05,0}
\definecolor{mywhite}{rgb}{1,1,1}
\definecolor{myred}{rgb}{0.7,0.13,0.13}

 \def\ee{\mathord{\rm e}}
 
 \def\ii{\mathord{\rm i}}

\def\half{\textstyle\frac{1}{2}}

\renewcommand{\ii}{{\rm i}}
\renewcommand{\ee}{{\rm e}}

 \newcommand{\ket}[1]{|#1\rangle}
 \newcommand{\bra}[1]{\langle #1|}

\begin{document}

\title[Short Title]{Dynamical delocalization of Majorana edge states by sweeping across a quantum critical point}

\author{A. Bermudez$^1$,  L. Amico$^{2}$, and M. A. Martin-Delgado$^{1}$}

\affiliation{$^1$Departamento de F\'{i}sica Te\'orica I,
Universidad Complutense, 28040 Madrid, Spain \\
$^2$MATIS-INFM  $\&$ Dipartimento di Metodologie Fisiche Chimiche(DMFCI), 
Universit\`{a} di Catania, viale A. Doria 6, 95125 Catania, Italy}

\begin{abstract}
We study the adiabatic dynamics of Majorana fermions  across a quantum phase transition. We show that the Kibble-Zurek scaling, which describes the density of bulk defects produced during the critical point crossing,  is not valid for  edge Majorana fermions. Therefore, the dynamics governing an edge state quench is non-universal and depends on the topological features of the system. Besides, we show that the localization of Majorana fermions is a necessary ingredient to guaranty  robustness against defect production.

\end{abstract}
\date{\today}
\maketitle

\section{Introduction}

When certain control parameter is varied, quantum mechanical fluctuations may drive a critical change of the system ground state~\cite{sachdev}. Even though such  phase transitions occur at zero temperature,  they have a profound influence on phenomena as diverse as  high-T$_{\text{c}}$ superconductivity~\cite{kivelson}, magnetism, or  quantum Hall effects.  The potential in both fundamental and applied research has stimulated an outgrowing interest in the physical community to study the effects  of the dynamical crossing of quantum critical points~\cite{Kibble,Zurek,Zurek_2,KZ_classical,KZ_classical_2,KZ_quant,KZ_ising_an, KZ_scaling_alt,KZ_landau_zener,KZ_xy_model,KZ_xy_bfield,KZ_spinor_cond,KZ_ferro_bec,KZ_bose_hubbard,KZ_ring_ol,KZ_exp,KZ_non_linear_quench, KZ_non_linear_quench_opt,KZ_multicritical_point,tri_critical,KZ_across_critical_region,KZ_across_critical_region_2,KZ_along_critical_line,KZ_classical_noise,KZ_environment, KZ_environment_2,KZ_environment_3,KZ_edge_states}.  At such points, the correlation length of the system diverges, and the characteristic energy gap between the ground state and the lowest lying excitations vanishes.  Accordingly, adiabatic evolution is precluded, and any dynamical quench across the critical point is accompanied by a production of  excitations in the system. As a result, the final state will only be partially ordered, displaying a non-vanishing density of defects imported from the quantum disordered into the ordered phases. Such density of defects is uniquely determined by the  universality class of the system, and  can be accurately described by the  so-called
Kibble-Zurek scaling (KZ)~\cite{Kibble,Zurek}. 

The KZ scaling was formerly proposed as the mechanism underlying topological defect production in a cosmological scenario~\cite{Kibble,Zurek,Zurek_2},  or in classical phase transitions occurring at finite temperature\cite{KZ_classical,KZ_classical_2}.  
The extension of the KZ mechanism to the quantum domain was proposed in~\cite{KZ_quant}, and numerically confirmed in  the transverse Ising model, a  cornerstone in the theory of quantum phase transitions. This  result paved the way to a considerable amount of works dealing with defect production in the zero-temperature regime~\cite{KZ_ising_an, KZ_scaling_alt,KZ_landau_zener,KZ_xy_model,KZ_xy_bfield,KZ_spinor_cond,KZ_ferro_bec,KZ_bose_hubbard,KZ_ring_ol}, which showed that the KZ scenario also  holds  for quantum phase transitions (see~\cite{KZ_exp} for an experiment). Modifications to the characteristic scaling laws have been studied  in order to account  for  non-linear quenches~\cite{KZ_non_linear_quench, KZ_non_linear_quench_opt}, crossing of multi-critical points~\cite{KZ_multicritical_point} (see ~\cite{tri_critical} for a recent proposal on the implementation of  tri-critical points in trapped ions), quenches across a critical region~\cite{KZ_across_critical_region,KZ_across_critical_region_2},  or along a gapless line~\cite{KZ_along_critical_line}. The effect of classical noise on the driving parameter has also been studied in~\cite{KZ_classical_noise}. In~\cite{KZ_environment, KZ_environment_2,KZ_environment_3}, it was shown that the density of defects scales with the rate of the quench even at finite temperatures in a regime of  incipient  criticality. 

Recently, we evidenced that the ground state  topology  of the system modifies the quench dynamics in a subtle manner~\cite{KZ_edge_states}. In particular, we showed how single fermions bound to the edges of the Creutz ladder~\cite{creutz} induce an anomalous defect production as the magnetic flux is swept across a quantum critical point.  In this paper, we reinforce the scenario, by providing a further evidence  of anomalous defect production in  the adiabatic dynamics of edge states. We consider a set of Majorana fermions (i.e. real fermions) arranged in a one-dimensional chain~\cite{kitaev_majorana_chain,kitaev_les_houches}.  We shall describe the deviations of the standard KZ scaling, and identify the necessary ingredients to obtain robustness of edge states even in a critical region by resorting to the experience gained with the Creutz ladder~\cite{KZ_edge_states}. The sensitivity  to boundary conditions generically indicates that a non trivial topological order is encoded into the system. This is specified by the presence  of states localized at the boundaries of the system, the so-called edge states~\cite{wen_book}.   Note also that topological edge states arise naturally in a wide variety of systems, such as one-dimensional spin models~\cite{aklt}, the integer and fractional quantum Hall effects (QHE)~\cite{ iqh_edges, fqh_edges}, and have been experimentally realized on topological insulators~\cite{qshi_experiment,3d_ti_experiment,3d_ti_experiment2}.

 The subtle difference between trivial and topological insulators (TI)  is rooted on the non-trivial topological features of the bulk bands of the latter, such as non-vanishing Chern numbers in the integer QHE~\cite{TKNN}. TI  represent an intriguing state of matter, where edge-transport exists even in the presence of an insulating bulk gap~\cite{top_insulator_nv}. These edge states are usually localized in the interface that separates two topologically distinct insulators, the simplest example being the interface between an IQHE sample and the vacuum~\cite{iqh_edges}. Besides, they connect the conduction and valence bands and allow for conduction along the interface. Other interesting TI are the anomalous half-integer QHE in the honeycomb~\cite{hiqh_honeycomb} or square lattices~\cite{hiqh_square}, and those ones characterized by $\mathbb{Z}_2$ topological invariants~\cite {qsh_insulator_graphene, qsh_insulator_z2,qshi_semicond,3d_top_insulator,3d_top_insulator_2,3d_top_insulator_3} which lead to the quantum spin Hall effect. Depending on the symmetries of the Hamiltonian and the dimension of the system, TI can be classified in a periodic table ~\cite{ti_classification_1,ti_classification_2}. Accordingly, one of the simplest TI is the one-dimensional chain of Majorana fermions~\cite{kitaev_majorana_chain}, where edge states correspond to bound Majorana fermions localized at the chain ends. The possibility to isolate Majorana fermions~\cite{majorana} represents an important step towards topological quantum computation~\cite{kitaev_tqc,review_tqc}, and can in principle be achieved in the vortex core of two-dimensional $p+\ii p$ superconductors~\cite{read_green}. A recent proposal is to create, fuse and transport Majoranas by the proximity effects in tri-junctions of superconductor-topological insulator-superconductor \cite{bound_majoranas_ti,bound_majoranas_detection}. Therefore, the results on the quench dynamics of the idealized Majorana chain discussed in this work might have applications in the more realistic scenario described in~\cite{bound_majoranas_ti,bound_majoranas_detection,bound_majoranas_teleportation,bound_majoranas_teleportation_2, lee, nagaosa,lee_2}.

This paper is organized as follows. In Sec.~\ref{maj_chain} we describe the nature of Majorana edge states in a one-dimensional lattice system with broken U(1) gauge symmetry. In Sec.~\ref{quench_KZ}, we present the results for the quench dynamics of the Majorana chain across a quantum phase transition. In particular, we show that bulk states fulfill the Kibble-Zurek prediction, whereas topological edge state  provide a clear anomaly. Finally, we present the conclusions of this work in the light of topological band insulators in Sec.~\ref{conclusions}. In Appendix~\ref{bulk_app}, we describe the main properties of bulk states in the periodic chain.

\section{The Majorana chain}
\label{maj_chain}

In this section, we review the properties of the Majorana chain~\cite{kitaev_majorana_chain}, a fermionic one-dimensional system that breaks time-reversal and $\text{U}(1)$ gauge symmetries, and presents Majorana fermions bound to its edges. We present a detailed description of the different phases where edge states occur, and discuss the nature of the quantum phase transition connecting the aforementioned topological phases. A discussion on the bulk spectrum is presented in Appendix~\ref{bulk_app}.

Let us consider a system of spinless fermions hopping in a
one-dimensional chain according to the following Hamiltonian
\begin{equation}
\label{fermion_hamiltonian}
H=\sum_j\left(-wa^{\dagger}_ja_{j+1}+\Delta a_ja_{j+1}-\textstyle{\frac{\mu}{2}}a_j^{\dagger}a_j+\text{h.c}\right),
\end{equation}
where $a_j$ $(a_j^{\dagger})$ represent spinless fermionic annihilation (creation) operators satisfying  canonical anticommutation relations $\{a_j,a_k^{\dagger}\}=a_ja_k^{\dagger}+a_k^{\dagger}a_j=\delta_{jk}$. As a second quantized Hamiltonian, Eq.~\eqref{fermion_hamiltonian} describes a mean-field Bardeen-Cooper-Schrieffer (BCS)  superconductor~\cite{tinkham_book}, where $w$ is the hopping amplitude, $\Delta=|\Delta|\ee^{\ii\theta}$ stands for the superconducting gap, and $\mu$ stands for the chemical potential. Due to the U(1) symmetry breaking inherent to superconductors (i.e. absence of fermion number conservation), it is more appropriate to describe the system in terms of Majorana fermions 
\begin{equation}
\begin{split}
\label{majorana_op}
c_{2j-1}=\textstyle{\frac{1}{\sqrt{2}}}\left(\ee^{-\frac{\ii\theta}{2}}a^{\dagger}_j+\ee^{\frac{\ii\theta}{2}}a_j\right),\\\hspace{0.5ex}c_{2j}=\textstyle{\frac{\ii}{\sqrt{2}}}\left(\ee^{-\frac{\ii\theta}{2}}a^{\dagger}_j-\ee^{\frac{\ii\theta}{2}}a_j\right),
\end{split}
\end{equation}
which are hermitian operators $c_{2j-1}^{\dagger}=c_{2j-1},c_{2j}^{\dagger}=c_{2j}$ satisfying the Majorana fermionic algebra $\{c_{j},c_k\}=\delta_{jk}$.  Although Majorana excitations are commonly paired to constitute a standard fermion, very recent research activity proposes to isolate and manipulate them, being  topological quantum computation the major motivation for doing so~\cite{kitaev_tqc,review_tqc}. Here,  $\nu=\textstyle{\frac{5}{2}}$ fractional quantum Hall effect~\cite{fqh_1,fqh_2,fqh_3}  or time-reversal topological insulators~\cite{bound_majoranas_ti} are the most promising candidates. 

\begin{figure}[!hbp]

\centering
\begin{overpic}[width=6.50 cm]{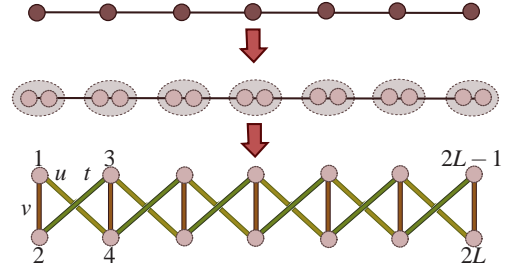}
\put(2,7){$v$}
\put(9.0,14){$u$}
\put(15,14){$t$}
\put(4.5,17){$1$}
\put(4.5,-3){$2$}
\put(19,17){$3$}
\put(19,-3){$4$}
\put(88,17){$2L-1$}
\put(92,-3){$2L$}
\end{overpic}

\caption{Scheme of the Majorana chain represented as a virtual ladder. The initial fermionic chain is mapped onto the Majorana fermions by means of Eq.~\eqref{majorana_op}, which then can be represented in a virtual ladder, where the $j$-th rung contains two Majoranas $c_{2j-1},c_{2j}$ and corresponds to the $j$-th site of the original chain. In the ladder scheme, the Majorana couplings in Eq.~\eqref{majorana_hamiltonian} become simple vertical and diagonal links.}
 \label{scheme_ladder}
\end{figure}

In order to understand how the Majorana chain  supports unpaired Majorana fermions at the boundaries, we need to rewrite the Hamiltonian of Eq.~\eqref{fermion_hamiltonian} in the Majorana picture
\begin{equation}
\label{majorana_hamiltonian}
H=\ii\sum_j\left(tc_{2j}c_{2j+1}+uc_{2j-1}c_{2j+2}+vc_{2j-1}c_{2j}\right),
\end{equation}
where $t=w+|\Delta|$, $u=-w+|\Delta|$, and $v=-\mu$ represent the different couplings between Majoranas, which might be interpreted as simple links in the ladder configuration of fig.~\ref{scheme_ladder}.  

 We shall be interested in the following regimes:

\vspace{0.75ex}
{\it a) Left-handed regime:} This regime occurs at $(t,u,v)=(0,2|\Delta|,0)$, and presents the following  zero-energy edge states, completely localized within the boundary rungs 
\begin{equation}
\label{left_edge_states}
\ket{\text{l}}_l=
\left |^{\circ} _ {\bullet}{ }^{\circ} _ {\circ}\cdots ^{\circ} _ {\circ}{}^{\circ} _ {\circ}\right \rangle =c_2\ket{\Omega},\hspace{0.5ex}\ket{\text{r}}_l=\left |^{\circ} _ {\circ}{ }^{\circ} _ {\circ}\cdots ^{\circ} _ {\circ}{}^{\bullet} _ {\circ}\right \rangle =c_{2L-1}\ket{\Omega},
\end{equation}
where $\ket{\Omega}$ stands for the Bogoliubov vacuum (see Eq.~\eqref{bog_vac} in Appendix~\ref{bulk_app}). As represented in fig.~\ref{left_handed_edges}, this is the only possibility to obtain localized Majoranas, since the remaining bulk eigenstates consist of paired of Majoranas with energies $\epsilon_{j\pm}=\pm 2|\Delta|$, and eigenstates
\begin{equation}
\ket{\epsilon_{j\pm}}_l=\left |^{\circ} _ {\circ}{ }^{\circ} _ {\circ}\cdots ^{\circ} _ {\circ}{}^{\bullet} _ {\circ}{}^{\circ} _ {\bullet}{}^{\circ} _ {\circ}\cdots ^{\circ} _ {\circ}{}^{\circ} _ {\circ}\right \rangle =\textstyle{\frac{1}{\sqrt{2}}}(\pm\ii c_{2j-1}+c_{2j+2})\ket{\Omega}.
\end{equation}
The bulk solution consist of a flat band of  plaquettes (i.e. two rungs) that contain a couple of Majoranas dimerized into a standard complex fermion. 

\vspace{0.75ex}
{\it b) Right-handed regime:} In this regime $(t,u,v)=(2|\Delta|,0,0)$, we obtain the following  zero-energy edge states completely localized within the boundary rungs 
\begin{equation}
\ket{\text{l}}_r=\left |^{\bullet} _ {\circ}{ }^{\circ} _ {\circ}\cdots ^{\circ} _ {\circ}{}^{\circ} _ {\circ}\right \rangle =c_1\ket{\Omega},\hspace{0.5ex}\ket{\text{r}}_r=\left |^{\circ} _ {\circ}{ }^{\circ} _ {\circ}\cdots ^{\circ} _ {\circ}{}^{\circ} _ {\bullet}\right \rangle =c_{2L}\ket{\Omega}.
\end{equation}
In fig.~\ref{right_handed_edges}, we represent the localized Majoranas, and the bulk plaquette-like eigenstates with energies $\epsilon_{j\pm}=\pm 2|\Delta|$ become
\begin{equation}
\ket{ \epsilon_{j\pm}}_r=\left |^{\circ} _ {\circ}{ }^{\circ} _ {\circ}\cdots ^{\circ} _ {\circ} {}^{\circ} _ {\bullet}{}^{\bullet} _ {\circ} {}^{\circ} _ {\circ}\cdots ^{\circ} _ {\circ}{}^{\circ} _ {\circ}\right \rangle=\textstyle{\frac{1}{\sqrt{2}}}(\pm\ii c_{2j}+c_{2j+1})\ket{\Omega}.
\end{equation}
Note how the bulk dimers containing a couple of paired Majoranas have been tilted with respect to the left-handed regime in fig.~\ref{left_handed_edges} 

\vspace{0.75ex}
{\it c) Topologically-trivial regime:} In the regime $|t|,|u| \ll |v|$, every Majorana is paired up with its rung partner, and form plaquette-like eigenstates in flat bands $\epsilon_{j\pm}=\pm \mu$
\begin{equation}
\ket{\epsilon_{j\pm}}=\left |^{\circ} _ {\circ}{ }^{\circ} _ {\circ}\cdots ^{\circ} _ {\circ}{}^{\bullet} _ {\bullet}{}^{\circ} _ {\circ}\cdots ^{\circ} _ {\circ}{}^{\circ} _ {\circ}\right \rangle \textstyle{\frac{1}{\sqrt{2}}}(\mp\ii c_{2j-1}+c_{2j})\ket{\Omega}.
\end{equation}
Therefore, in this trivial regime edge states are forbidden (see fig.~\ref{no_edges}).

\begin{figure}[!hbp]
\centering
\subfigure[\hspace{1ex}Left-handed regime ]{
\begin{overpic}[width=5.5 cm]{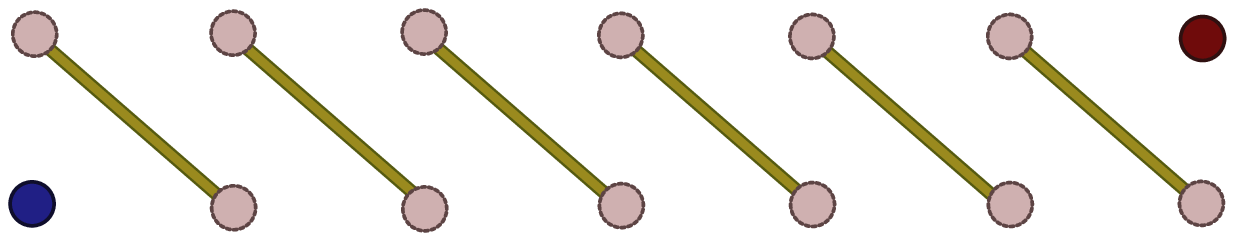} \label{left_handed_edges}
\put(0,6){$c_2$}
\put(95,11){$c_{2L-1}$}
\end{overpic}
}
\subfigure[\hspace{1ex}Right-handed regime]{
\begin{overpic}[width=5.50 cm]{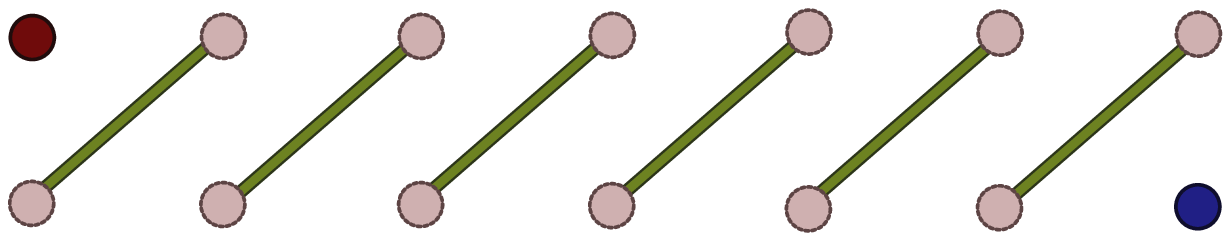} \label{right_handed_edges}
\put(0,11){$c_1$}
\put(95,6){$c_{2L}$}
\end{overpic}
}
\subfigure[\hspace{1ex}Topologically-trivial regime]{
\begin{overpic}[width=5.50 cm]{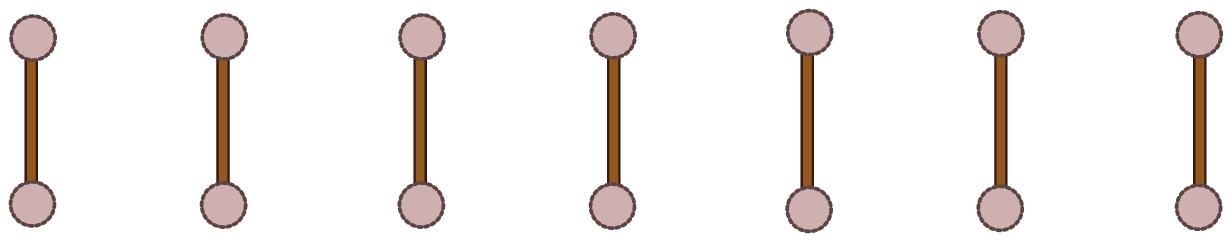} \label{no_edges}
\end{overpic}
}
\caption{Scheme of the unpaired Majorana fermions bound at the edges, and the dimerized Majoranas at the bulk. (a) In the left-handed regime, only the left-tilted diagonal links survive, and thus a couple of unpaired Majoranas appear at sites $2,2L-1$ of the ladder. (b) In the right-handed regime, only the right-tilted diagonal links survive, and thus a couple of unpaired Majoranas appear at sites $1,2L$. (c) In the topologically-trivial regime, every Majorana fermion is paired within a single rung, and thus no edge can exist.}
\end{figure}

We have thus shown that the Majorana chain Hamiltonian introduced in Eq.~\eqref{majorana_op} entails different phases where edge states are pinned to the system boundary, but also trivial phases where unpaired Majoranas cannot exist. We comment  that the Hamiltonian in Eq.~\eqref{fermion_hamiltonian}  could be realized with junctions involving s-wave superconductors and topological insulators~\cite{bound_majoranas_ti}. In particular, these different phases can be engineered in  superconductor-topological insulator tri-junctions, where the key parameter that determines the existence of unpaired Majoranas is the relative phase of the superconductors. In the following section, we describe the quantum phase transitions between the phases introduced above, and study how the edge and bulk states behave when the system is driven across a quantum critical point.

\section{Quench between edge states}
\label{quench_KZ}

\begin{figure}[!hbp] 

\centering
\begin{overpic}[width=6.0 cm]{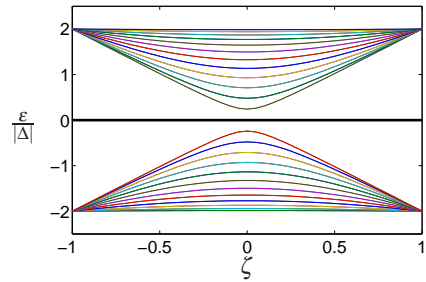} 
\put(-1,30){$\frac{\epsilon}{|\Delta|}$}
\put(50,-2){$\zeta$}
\end{overpic}
\caption{Energy spectrum $\epsilon/|\Delta|$ of the open Majorana chain for $t=0$, and relative parameter $\zeta=w/|\Delta|\in[-1,1]$.}
\label{xi_energies}
\end{figure}

In this section, we study the quantum phase transition between the left- and right-handed phases (see {\it a)} and {\it b)} above). Such a quantum phase transition can be characterized by $t=0$ and a relative parameter $\zeta=w/|\Delta|$, so that the left-handed regime is realized at $\zeta=-1$ and the right-handed at $\zeta=1$. In fig.~\ref{xi_energies}, we depict the energy spectrum of the Majorana Hamiltonian in Eq.~\eqref{majorana_hamiltonian}  for $\zeta\in[-1,1]$, which clearly shows a quantum critical point at $\zeta_c=0$ where the energy gap vanishes. Note also how two zero-energy modes, corresponding to the unpaired Majoranas introduced above,  are present for arbitrary $\zeta$. Let us remark here that such edge modes are absent in the periodic chain (see Eq.~\eqref{bulk_energies} and fig.  \ref{xi_energies_per} in Appendix~\ref{bulk_app}).

The critical properties of this quantum phase transition are characterized by the critical exponents $\nu=z=1$,
which describe how the coherence length $\xi_l$ diverges and the energy gap $\Delta\epsilon$ vanishes at criticality $\xi_l\sim{\Delta\epsilon}^{-1}\sim|\zeta-\zeta_c|^{-1}$.
We shall consider the adiabatic dynamics of a state initially localized within the bulk or edge of the system for $w(t_0)=-|\Delta|$.
At the end of the quench $w(t_{\text{f}})=-|\Delta|$, the system 
shall be found in a state $\ket{\Psi(t_{\text{f}})}$ given by the superposition of the adiabatically-connected state, which is still localized, and delocalized defects which have been excited close to the 
critical point. Since we are concerned with single-particle edge- or bulk-states, we shall refer to the probability to create a defect $P_{\text{def}}=\sum_{E>0}|\langle E|\Psi(t_{f}\rangle|^2$, with $\bra{E}$ being the eigenvectors of 
the Hamiltonian. Interestingly enough, for a linear adiabatic quench $\zeta(t)=-1+v_Qt$ with rate $v_Q\ll 1$ and lasting for $t\in[0,2/v_Q]$, the Kibble-Zurek mechanism predicts that such probability scales as
\begin{equation}
P^{\text{KZ}}_{\text{def}}\sim v_Q^{\frac{d\nu}{z\nu+1}},
\end{equation}
where $d$ is the dimension of the system~\cite{Kibble,Zurek}. In the present phase transition, the KZ mechanism predicts a scaling $P_{\text{def}}^{\text{KZ}}\sim\sqrt{v_Q}$. In the following sections, we shall confront the KZ prediction with the exact dynamics of the Majorana chain.

\subsection{Quench dynamics in the periodic  chain}
\label{bulk_quench_app}

\begin{figure}[!hbp]
\centering
\begin{overpic}[width=7.50 cm]{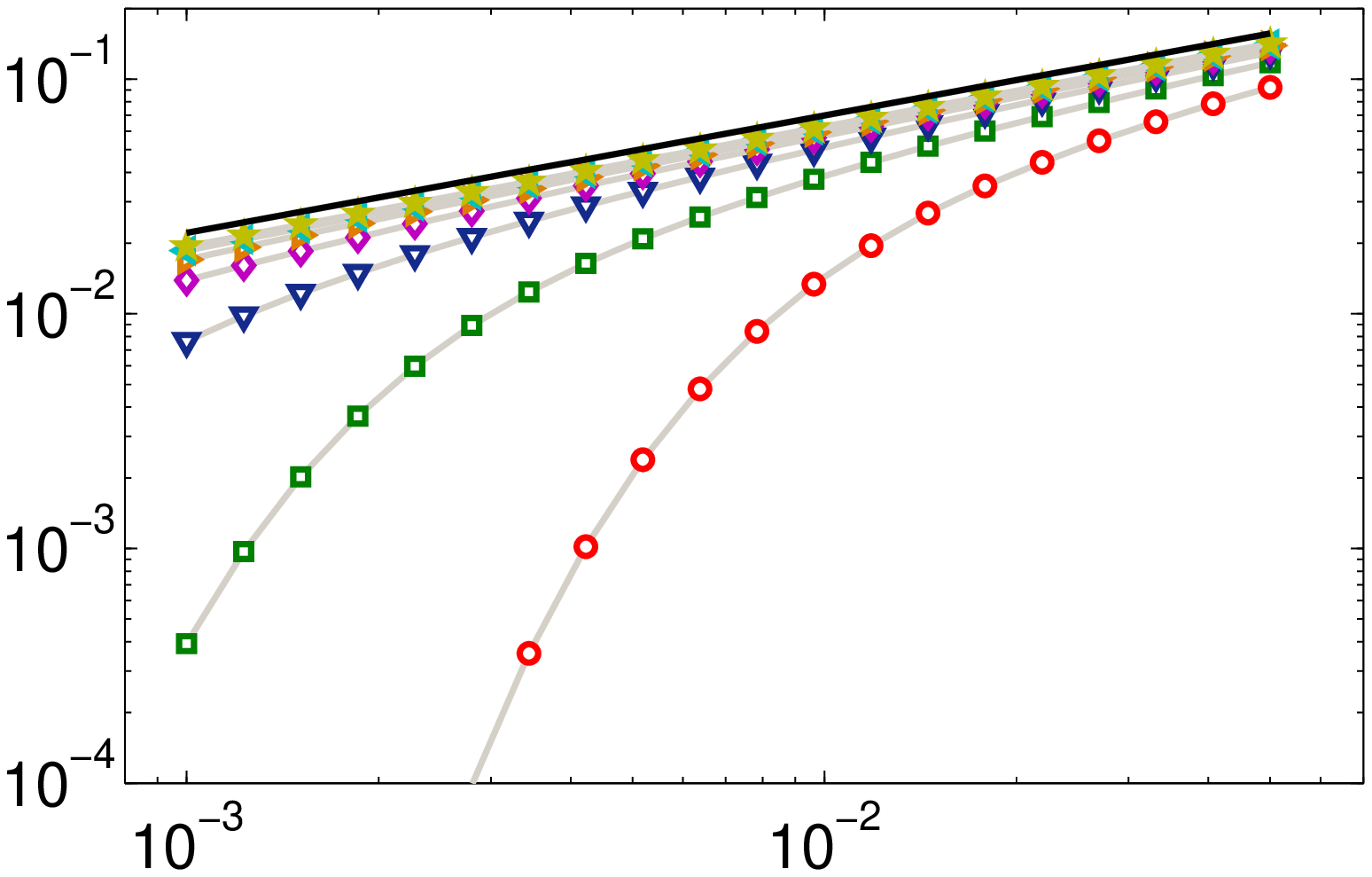}
\put(-2,42){$\log P_{\text{def}}$}
\put(40,3){$\log v_Q$}
\put(36,50){$P_{\text{def}}\sim\sqrt{v_Q}$}
\end{overpic}
\caption{Scaling of the defects produced by adiabatically crossing the $\zeta_c=0$ critical point for the periodic lattice with number sites $L\in\{40,80,160,320,640,1280,2560\}$.}
 \label{bulk_kz_periodic}
\end{figure}

In Appendix~\ref{bulk_app}, we describe the energy spectrum of the translationally-invariant Majorana chain, and show how the single-particle energy levels can be expressed as excitations over the Bogoliubov vacuum (i.e. $\gamma_{q,\pm}^{\dagger}\ket{\Omega}$). Here, we  study the quench dynamics of the periodic chain when the parameter $\zeta$ is adiabatically modified across the corresponding critical point. In the translationally-invariant case, no edge states exist, and the only localized state in the negative energy band $E=-2|\Delta|$ at $\zeta=-1$, is a plaquette-like fermion 
\begin{equation}
\ket{\Psi(0)}=\textstyle{\frac{1}{\sqrt{2}}}(-\ii c_{2j-1}+c_{2j+2})\ket{\Omega}=\sum_q\Psi_q^{\dagger}\boldsymbol{c}_q(t_0)\ket{\Omega},
\end{equation}
where $\boldsymbol{c}_q(t_0)=\frac{\ii\ee^{+\ii qj}}{2\sqrt{L}}(\ee^{-\frac{\ii\theta}{2}}(-1+\ee^{iq}), -\ee^{+\frac{\ii\theta}{2}}(1+\ee^{iq}))$ determines such initial state in momentum Nambu-representation. Remarkably, the whole quench dynamics can be expressed as a collection of uncoupled two-level systems, each associated to a Nambu spinor $\Psi_q$. Therefore, the dynamics under $\zeta(t)=-1+v_Q t$ entails an ensemble of two-level Landau-Zener processes~\cite{KZ_landau_zener, LZ, LZ_bis}, and the final excitation probability (compare to Eq.~\eqref{ex_density}) can be simply calculated as
\begin{equation}
P_{\text{def}}=\sum_q |\boldsymbol{c}_{q+}^{\zeta=1}\cdot \boldsymbol{c}_q(t_f)|^2,
\end{equation}
where $ \boldsymbol{c}_q(t_f)$ is the final state after the quench evolution, and $\boldsymbol{c}_{q+}^{\zeta=1}$ correspond to the positive-energy eigenstates at $\zeta=+1$. The evaluation of the excitation probabilities is simpler in this translationally invariant case, and we can thus treat larger chains and slower quenches. In fig.~\ref{bulk_kz_periodic}, we observe that the scaling fully agrees with the KZ-prediction $P_{\text{def}}\sim\sqrt{v_Q}$. 
 
In this section we have considered a single-particle initial state localized within a rung of the Majorana ladder.
 Let us note that the KZ-scaling  also holds for an initial ground state, which consists of all the negative-energy modes being occupied. For
 such state, one usually refers to the density of defects $n_{\text{def}}^{\text{KZ}}\sim\sqrt{v_q}$, which is finite in the thermodynamical limit.
 However, since we are interested in the comparison of initial single-particle bulk-, or edge-states, we shall refer to the probability to create defects $P_{\text{def}}$
 rather than to a thermodynamical density.

\subsection{Quench dynamics in the open  chain}
\label{open_quench_app}
\begin{figure}[!hbp]
\centering
\subfigure[\hspace{1ex}Adiabatically connected bulk plaquettes ]{
\begin{overpic}[width=7.0 cm]{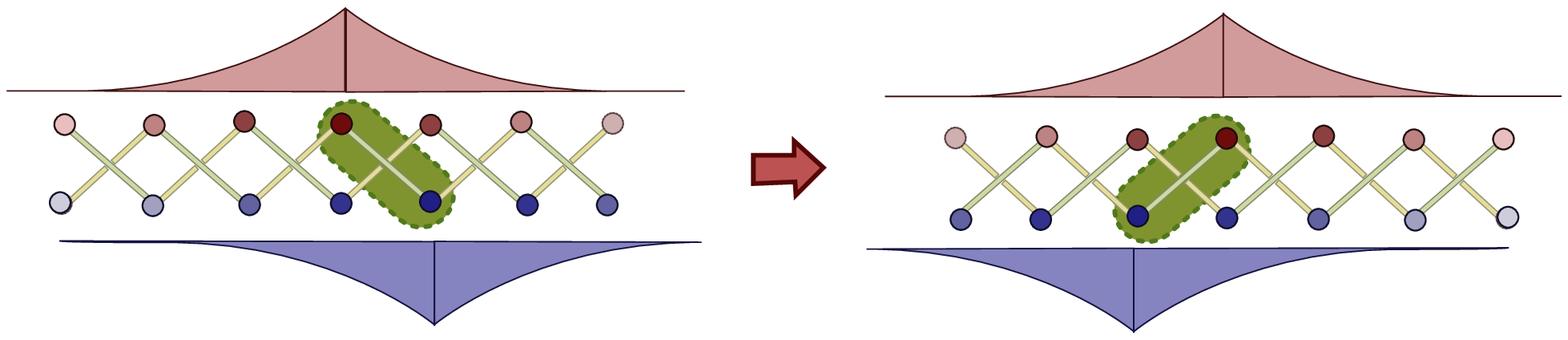} \label{bulk_adiab}
\end{overpic}
}
\subfigure[\hspace{1ex}Adiabatically connected edge Majoranas]{
\begin{overpic}[width=7.0 cm]{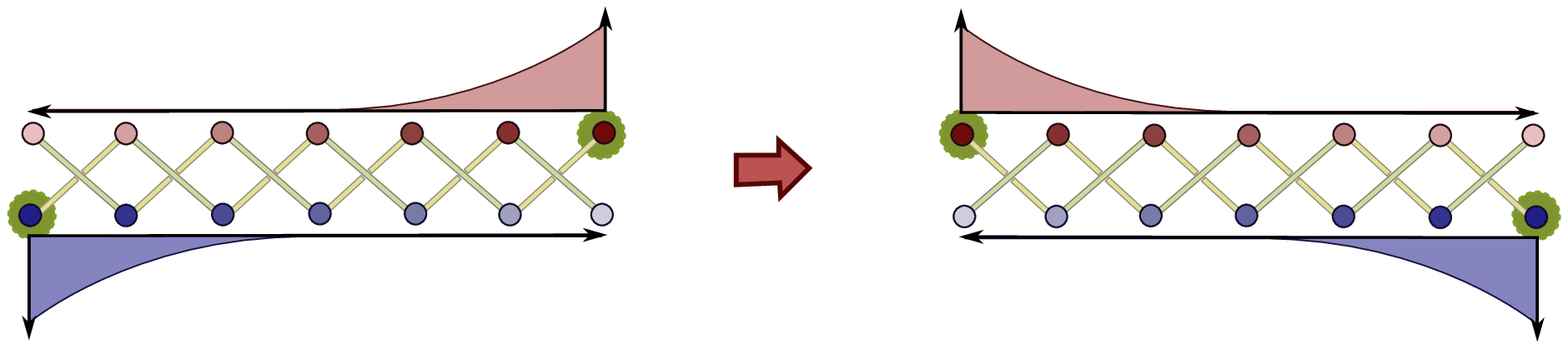} \label{edge_adiab}
\end{overpic}
}

\caption{Adiabatically connected states for $\zeta\in[-1,1]$ (a) In the chain bulk, left- and right-tilted plaquettes with energies $\epsilon=-2|\Delta|$ are connected $\ket{ \epsilon_{j-}}_l\rightarrow\ket{ \epsilon_{j-}}_r$. (b) In the chain edges, unpaired Majorana fermions at different edges are connected $\ket{\text{l}}_l \rightarrow \ket{\text{r}}_r$, and $\ket{\text{r}}_l \rightarrow \ket{\text{l}}_r$~\cite{comment1}.  }
\end{figure}

The objective of this section is to test the prediction for an initial plaquette-like state in the bulk, and for an initial edge Majorana fermion (see figs.~\ref{bulk_adiab} and~\ref{edge_adiab} for the adiabatic evolution of such initial states).

\begin{figure}[!hbp]
\centering
\subfigure[\hspace{1ex}Produced defects for  a bulk state quench ]{
\begin{overpic}[width=7.50 cm]{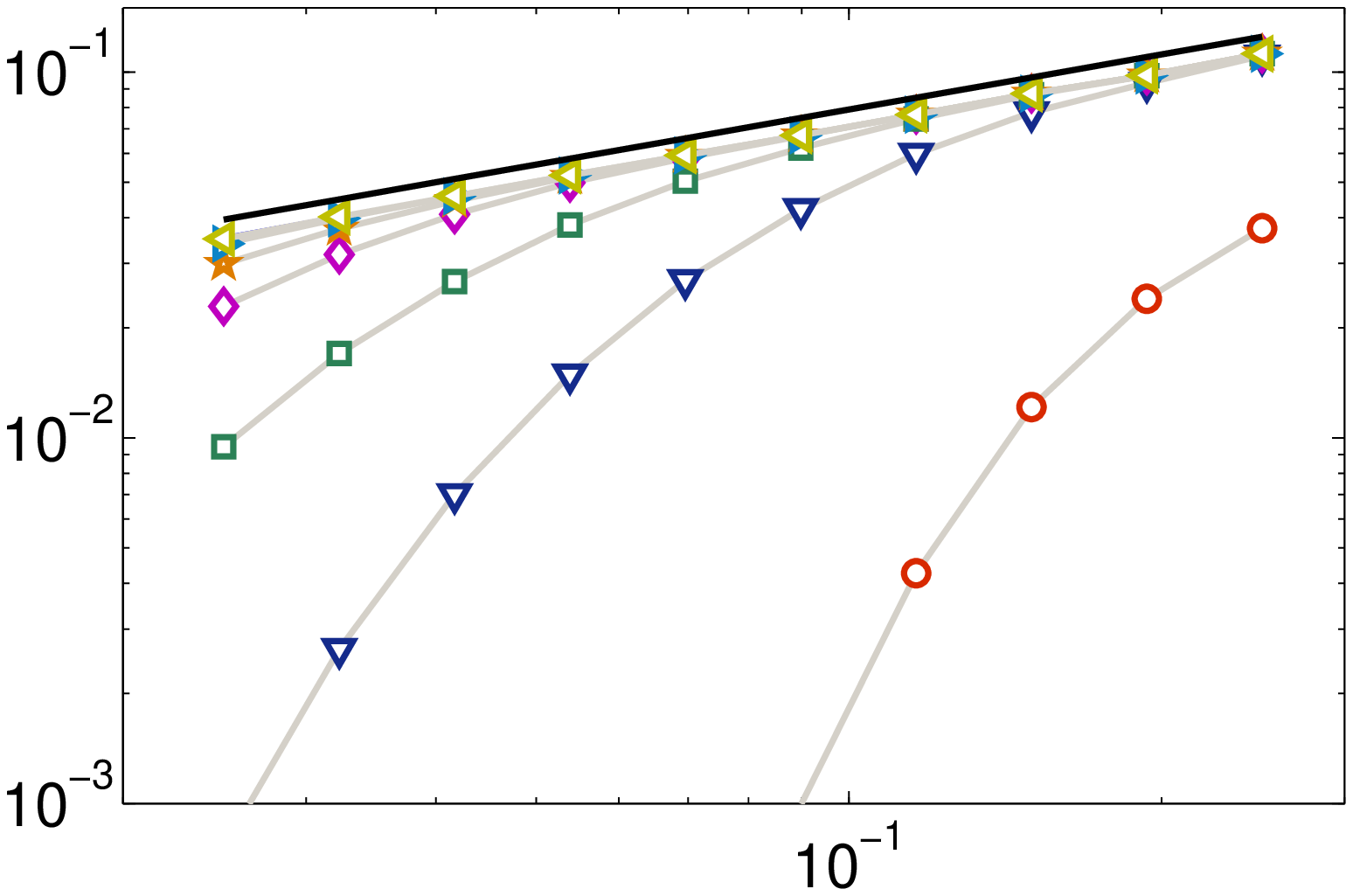} \label{bulk_kz}
\put(-1,40){$\log P_{\text{def}}$}
\put(45,3){$\log v_Q$}
\put(42,53){$P_{\text{def}}\sim\sqrt{v_Q}$}
\end{overpic}
}
\subfigure[\hspace{1ex}Produced defects for an edge state quench]{
\begin{overpic}[width=7.50 cm]{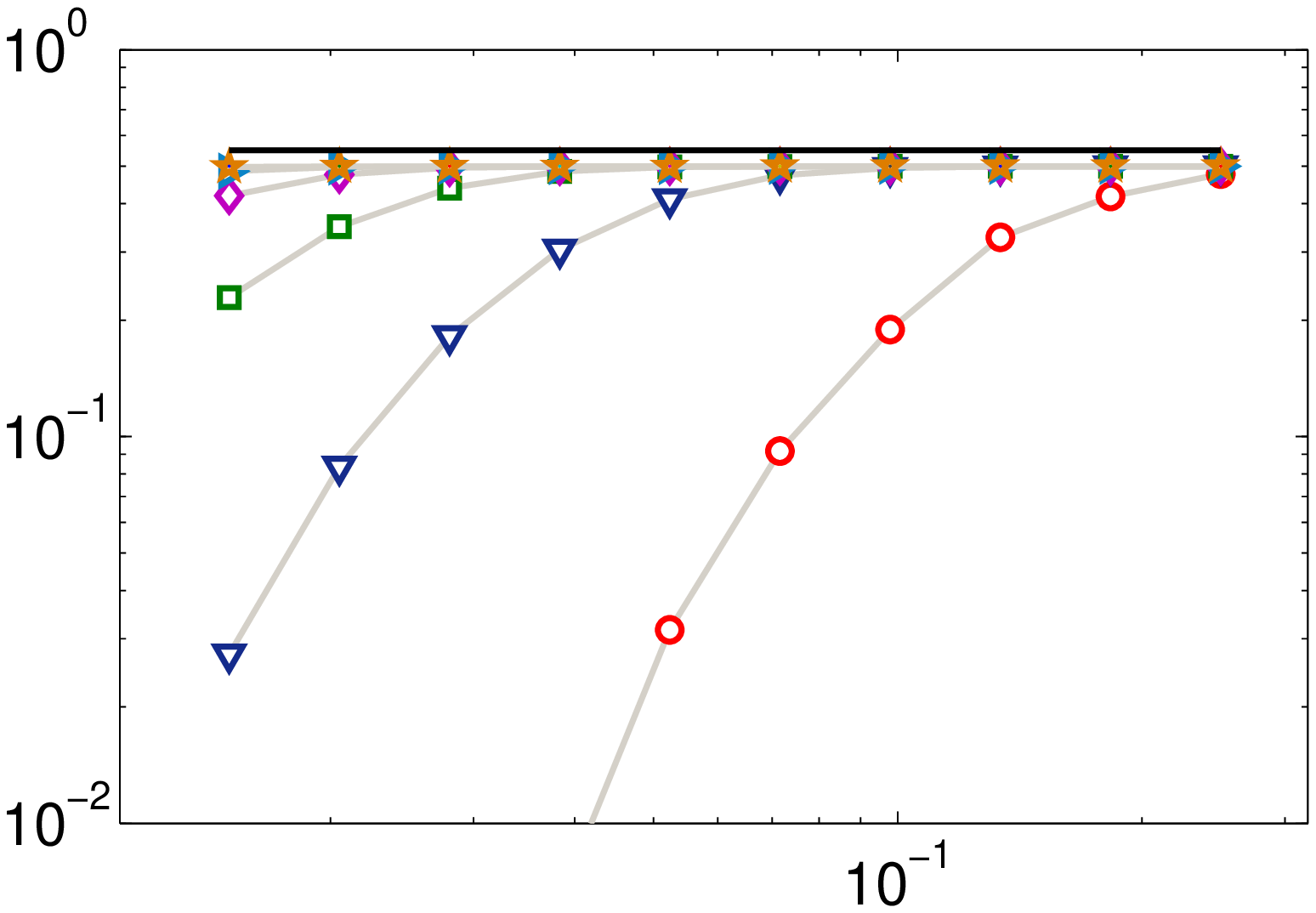} \label{edge_kz}
\put(-1,40){$\log P_{\text{def}}$}
\put(45,3){$\log v_Q$}
\put(42,53){$P_{\text{def}}\sim{v_Q}^0$}
\end{overpic}
}

\caption{Scaling of the defects produced by adiabatically crossing the $\zeta_c=0$ critical point for lattice with number sites $L\in\{10,20,30,40,50,60,70,80,160\}$. (a) Defects $P_{\text{def}}$ for an initial negative-energy plaquette localized at the bulk. (b) Defects $P_{\text{def}}$ for an initial zero-energy Majorana localized at the left edge of the sample.  }
\end{figure}

\vspace{0.75ex}
{\it Adiabatic quench for an initial plaquette:} In this case, the initial state  corresponds to a negative-energy $\epsilon=-2|\Delta|$ localized in the bulk $\ket{\Psi(0)}=\textstyle{\frac{1}{\sqrt{2}}}(-\ii c_{L-1}+c_{L+2})\ket{\Omega}$ for $\zeta=-1$.  We evaluate the defects produced close to the critical point as follows
\begin{equation}
\label{ex_density}
P_{\text{def}}=\sum_{\epsilon>0}|\langle \epsilon|\Psi(t_{\rm f})\rangle|^2
\end{equation}
Let us emphasize that the initial bulk fermion can be expressed as linear combination of every periodic eigenstate in the negative-energy band (see Eq.~\eqref{bog_op}). In particular, the modes responsible for the universal KZ-scaling (i.e. those around the gapless mode $q\sim0$) shall be populated, and one thus expects  the KZ prediction to hold. Indeed, the numerical results in fig.~\ref{bulk_kz} are in clear agreement with the predicted scaling $P_{\text{def}}^{\text{KZ}}\sim\sqrt{v_Q}$ at the thermodynamical limit $L\to\infty$ (see also fig.~\ref{bulk_kz_periodic} for the KZ-scaling in a periodic chain).

\vspace{0.75ex}
{\it Adiabatic quench for an initial edge Majorana:} In this case, the initial state  corresponds to a zero-energy Majorana fermion bound to the left edge $\ket{\Psi(0)}=\ket{\text{l}}_l=c_2\ket{\Omega}$ for $\zeta=-1$. Let us note that this state also crosses the critical point at the gapless mode energy, and thus one would expect a similar KZ-scaling to hold $P_{\text{def}}^{\text{KZ}}\sim\sqrt{v_Q}$. Note however, that the obtained scaling (fig.~\ref{edge_kz}) completely disagrees $P_{\text{def}}\sim (v_Q)^0$, and offers thus further evidence of the anomalous defect production for localized edge states~\cite{KZ_edge_states}.  The departure from the universal KZ-scaling is drastic for such Majorana fermions. Besides, it predicts that the edge state robustness is lost as the critical point is crossed (i.e. $P_{\text{def}}=\half$ both for positive- and negative-energy bands). The underlying reason for such a departure is that the edge state at the critical point can be expressed as a linear combination of the gapless modes $q=\{0,\pi\}$, and thus fuses into the bulk bands. Indeed, we  show in the next section that in the thermodynamical limit
\begin{equation}
\ket{\text{l}(\zeta_c)}_l\sim\left(c_2+c_6+\cdots+c_{2L-1}\right)\ket{\Omega}\sim\frac{1}{\sqrt{2}}\left(\gamma^{\dagger}_{0,+}+\gamma^{\dagger}_{-\pi,-}\right)\ket{\Omega},
\end{equation}
where $\gamma_{q,\pm}^{\dagger}$ are the fermionic creation operators of positive- and negative-energy solutions associated to the mode $q$ (see Eq.~\eqref{edge_critical} ). Therefore, the initially unpaired Majorana at the left edge, becomes completely delocalized along the upper chain of the rung (see fig.~\ref{edge_adiab})  at criticality $\zeta_c=0$. Being uniformly delocalized, it overlaps with bulk excitations with well-defined momenta $q=\{0,\pi\}$, and therefore completely fuses in the continuum bands. In this regard, the Majorana character of the initial state is completely lost across the quantum critical point, and one cannot further transport the Majorana to the right-hand side (i.e. $\ket{\text{l}}_l\nrightarrow \ket{\text{r}}_r$). 

Due to the finite number of modes coupled to the edge states, this 
	 anomalous dynamics might be viewed as a saturation effect, such as that one studied in
	 the XY model [26].  Nevertheless, we remark that the effect presented in this work traces 
	 back indeed to  the topological  order of the   system, since it only occurs for the unpaired 
	 Majorana edge states. In a trivial insulator, where in-gap edge states are simply absent, we 
	 should have restricted to bulk plaquette-like excitations which do  not lead to any  anomaly in  
	 KZ 	 scaling.


Let us now compare these results with the edge-state defect production in the Creutz model, a ladder of usual fermions pierced  by a perpendicular
 magnetic field~\cite{KZ_edge_states}. In this system, a quench of the magnetic flux drives the fermions across a quantum phase transition of  the same universality class $\nu=z=d=1$. Similarly to the Majorana chain, the defect production for initial bulk states fulfills $P_{\text{def}} \sim \sqrt{v_Q}$ in accordance to the KZ-scaling. This result emphasizes the fact that both models, being microscopically different, belong to the same universality class. Conversely, edge states in the Creutz ladder fulfill $P_{\text{def}} \sim v_Q^{1.35}$, which completely deviates from the KZ prediction, and from the defect production of a Majorana chain $P_{\text{def}} \sim v_Q^{0}$. This result gives further evidence of the conjecture on the non-universality of the edge state dynamics across a critical point, as  raised in~\cite{KZ_edge_states}. Moreover, the result on the Creutz ladder highlights the resilience of edge states due to a quantum interference phenomenon that allows them to remain well-localized at the boundaries of the lattice. Conversely, Majorana edge states become completely delocalized at the critical point, and this leads to a higher sensitivity at criticality $P_{\text{def}} \sim v_Q^{0}$. Therefore, we can draw the conclusion that edge state robustness is non-universal, and is intimately related to edge state localization.

\subsection{Adiabatic path for edge states}
\label{edge_app}

Here, we show how the unpaired Majoranas, initially localized at the chain edges, become completely delocalized over the whole chain when the critical point $\zeta_c=0$ is achieved. Besides, we show that such delocalized states overlap with the continuum modes in the thermodynamic limit $L\to\infty$, and thus justify the scaling $P_{\text{def}}\sim\half (v_Q)^0$. Remarkably, the zero-energy modes associated to unpaired Majoranas can be analytically obtained in the whole regime $-1\leq\zeta\leq0$, $\eta=0$, and read as follows
\begin{equation}
\begin{split}
\ket{\text{l}(\zeta)}&=\frac{1}{\mathcal{N}_l}\left(c_2+rc_6+\cdots+r^{\frac{L+1}{2}}c_{2L}\right)\ket{\Omega},\\
\ket{\text{r}(\zeta)}&=\frac{1}{\mathcal{N}_r}\left(c_{2L-1}+rc_{2L-5}+\cdots+r^{\frac{L+1}{2}}c_{1}\right)\ket{\Omega},
\end{split}
\end{equation}
where $r=(1+\zeta)/(1-\zeta)$, and $\mathcal{N}_{l,r}$ are normalization constants. Note that the exponential tail of the edge states fulfill $0\leq r\leq1$ in the regime of interest. In particular, at the initial point $\zeta=-1$, one can easily check that these edge states become completely localized at the edge rungs and coincide with Eqs.~\eqref{left_edge_states}. Conversely, at criticality $\zeta_c=0$, the Majoranas become uniformly delocalized along each chain forming the virtual rung
\begin{equation}
\begin{split}
\ket{\text{l}(\zeta_c)}&=\frac{1}{\sqrt{L}}(c_2+c_6+\cdots+c_{2L})\ket{\Omega},\\
\ket{\text{r}(\zeta_c)}&=\frac{1}{\sqrt{L}}(c_1+c_5+\cdots+c_{2L-1})\ket{\Omega}.
\end{split}
\end{equation}
Its precisely at this critical point where they overlap with the bulk gapless modes (see Eq.~\eqref {bog_op}), and thus
\begin{equation}
\label{edge_critical}
\begin{split}
\ket{\text{l}(\zeta_c)}&=\frac{1}{\sqrt{2}}\left(\gamma^{\dagger}_{0,+}+\gamma^{\dagger}_{-\pi,-}\right)\ket{\Omega},\\
\ket{\text{r}(\zeta_c)}&=\frac{1}{\sqrt{2}}\left(-\gamma^{\dagger}_{0,+}+\gamma^{\dagger}_{-\pi,-}\right)\ket{\Omega},
\end{split}
\end{equation}
Consequently, the edge states will not follow the adiabatic path that would connect them to the edge Majoranas at $\zeta=+1$, but rather fuse into the positive- and negative-energy bulk bands. This explains the anomalous scaling $P_{\text{def}}=\half$ for edge states (note also that the probability to decay into the negative energy band would be $P_{\text{neg}}=\half$).

\section{Conclusions}
\label{conclusions}

In this article, we have described the quench dynamics of a 1D-lattice model with unpaired Majorana fermions  bound to the boundaries~\cite{kitaev_majorana_chain}. The transport of Majorana fermions between the edges of the chain is hampered by the loss of adiabaticity at the quantum critical point. The main conclusion drawn from this work is two-fold: edge-state dynamics across a  critical point is not universal, and the robustness/fragility of the latter relies on their localization at criticality.

 We  have found that the scaling of  defects deviates from the usual Kibble-Zurek paradigm, and is non-universal. Indeed,  the Creutz ladder~\cite{KZ_edge_states} and the Majorana chain present different scalings within the same bulk universality class. With respect to our results on the Creutz ladder~\cite{KZ_edge_states}, we conclude that such deviations are due to  the localization dynamics of  edge states. We note, in particular,  that   Majorana edge states produce a peculiar scaling of excitations due to their interesting delocalization dynamics. At the quantum critical point, Majorana edge states 'fuse' with the bulk excitations, and thus become delocalized over the system. This should be contrasted with the phenomenology evidenced for the Creutz ladder, where  edge states stay localized for any value of the control parameter. Such a difference, in turn, traces back to the physical content of such states: for  the Creutz ladder, standard fermions are pinned to the ladder edges due to interference effects caused by  additional magnetic fields; for the Majorana chain, they are induced by the U(1) symmetry breaking term provided by a superconducting gap. Despite of this quantitative difference in the microscopic origin, we  note  that both  systems belong to the class of topological band insulators~\cite{ti_classification_1,ti_classification_2}. Therefore, we may conclude that quench dynamics of boundary states in topological insulators should provide additional anomalies in the defect production rate.

\vspace{1ex}
{\it Acknowledgments.} A.B. and M.A.M.D  acknowledge financial support 
from the Spanish MEC project FIS2006-04885, the 
project CAM-UCM/910758, the ESF Science Programme INSTANS 2005-2010. Additionally, A. B.
acknowledges support from a FPU MEC grant. We thank D. Patan\`e for useful and inspiring conversations.

\appendix

\section{Bulk spectrum of the Majorana chain}
\label{bulk_app}
\begin{figure}[!hbp]
\centering
\begin{overpic}[width=5.50 cm]{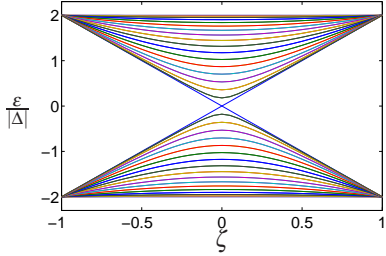}
\put(-1,30){$\frac{\epsilon}{|\Delta|}$}
\put(50,-2){$\zeta$}
\end{overpic}
\caption{Energy spectrum $\frac{\epsilon}{|\Delta|}$ of the periodic Majorana chain for $\eta=0$, and relative parameter $\zeta=w/|\Delta|\in[-1,1]$.}
 \label{xi_energies_per}
\end{figure}

In this Appendix, we describe the bulk spectrum of the Majorana chain for periodic boundary conditions ($a_{L+1}=a_1)$. Let us first point out  that the Hamiltonian in Eq.~\eqref{fermion_hamiltonian}, after  a standard Jordan-Wigner transformation~\cite{jordan_wigner}, describes  a set of localized spins governed by 
\begin{equation}
\begin{split}
H=\frac{-|\Delta|\cos\theta}{2}\sum_{j=1}^{N}&(\zeta\text{sec}\theta+1)\sigma^x_j\sigma^x_{j+1}+(\zeta\text{sec}\theta-1)\sigma^y_j\sigma^y_{j+1}\\
&+\tan\theta(\sigma^x_j \sigma^y_{j+1}+\sigma^y_j \sigma^x_{j+1}) +h(\sigma^z_j+1),
\end{split}
\label{modified-xy}
\end{equation}
where  $h=\mu/|\Delta|\cos\theta$  plays the role of an external transverse magnetic field, and the superconducting phase $\theta$ leads to the local lattice anisotropy in the exchange couplings, namely, $J_x=(\zeta\text{sec}\theta+1)$, and $J_y=(\zeta\text{sec}\theta-1)$, with $\zeta=w/|\Delta|$. Notice also the appearance of additional spin coupling terms $\sigma_{j}^x\sigma_{j+1}^y$, which are absent in standard anisotropic XY models~\cite{xy,xy_2}. Accordingly, a spin picture offers an alternative perspective on the U(1) symmetry-broken fermion system in Eq.~\eqref{fermion_hamiltonian}.
 In momentum space, the fermionic operators become $a_j=\textstyle{\frac{1}{\sqrt{L}}}\sum_{q}a_q\ee^{-\ii qj}$, where $q\in[-\pi,\pi]$ lies in the Brillouin zone and we have assumed unit lattice spacing. The initial Hamiltonian in Eq.~\eqref{fermion_hamiltonian} can be expressed as follows
\begin{equation}
H=\sum_q\Psi_q^{\dagger}H_q\Psi_q,\hspace{1ex}H_q=\left(\begin{array}{cc}\zeta_q & \Delta_q \\\Delta_q^* & -\zeta_q\end{array}\right),
\end{equation}
where we have introduced the Nambu spinor $\Psi_q=(a_q,a_{-q}^{\dagger})^{t}$, and $\zeta_q=-(w\cos q+\mu/2)$, $\Delta_q=\ii|\Delta|\ee^{-\ii\theta}\sin q$. The energy spectrum is obtained after a Bogoliubov transformation is performed, in a similar procedure as the  Bardeen-Cooper-Schrieffer (BCS) Hamiltonian is diagonalized~\cite{tinkham_book}. Transforming the fermion operators according to
\begin{equation}
\label{bog_op}
\gamma_{q+}=u_q^*a_q+v_qa_{-q}^{\dagger},\hspace{2ex}\gamma_{q-}=-v_q^*a_q+u_qa_{-q}^{\dagger},
\end{equation}
one obtains the energies $\epsilon_{q\pm}=\pm\epsilon_q=\pm\sqrt{\zeta_q^2+|\Delta_q|^2}$ corresponding to the positive- and negative-energy solutions $\gamma_{q\pm}$,
defined through the following  parameters
\begin{equation}
 u_q=\frac{1}{\sqrt{2}}\sqrt{1+\frac{\zeta_q}{\epsilon_q}}, \hspace{2ex} v_q=\frac{\ii\ee^{-\ii\theta}\text{sgn}(q)}{\sqrt{2}}\sqrt{1-\frac{\zeta_q}{\epsilon_q}}.
 \end{equation}
Besides, the solutions fulfill $\gamma_{-q\pm}=\gamma^{\dagger}_{q\mp}$, which allows us to restrict the Hamiltonian to half-Brillouin zone 
\begin{equation}
\label{periodic Hamiltonian}
H=\sum_{q>0}\left(2\epsilon_q\gamma_{q+}^{\dagger}\gamma_{q+}-2\epsilon_q\gamma_{q-}^{\dagger}\gamma_{q-}\right).
\end{equation}
Therefore, the final energy spectrum for a periodic Majorana chain is
\begin{equation}
\label{bulk_energies}
E_{q\pm}=\pm2\epsilon_q=\pm2|\Delta|\sqrt{(\eta+\zeta\cos q)^2+\sin^2q},
\end{equation}
where we have introduced the relative parameters $\zeta=w/|\Delta|,\eta=\mu/2|\Delta|$ used to study the quantum phase transitions in the open chain. In fig.~\ref{xi_energies_per}, we represent the energy spectrum as a function of $\zeta$ for $\eta=0$. We clearly identify a non-topological two-band insulator  since no mid-gap edge states occur at zero-energies.  This fact should be compared to the open ladder figs.~\ref{xi_energies}, where zero-energy  edge states localized at the interface between the topological insulator and vacuum arise. In these figures, it is also clear that the energy gap vanishes at the critical points $\zeta_c=0$, indicating thus a quantum phase transition. Furthermore, from the exact energy dispersion in Eq.~\eqref{bulk_energies}, one can show that the critical exponents of both transitions are $z=\nu=1$.

Let us finally comment on the system vacuum and single-particle states. One may define the Bogoliubov vacuum as a state satisfying $\gamma_{q\pm}\ket{\Omega}=0, \forall q\in \text{BZ}$. This vacuum can be constructed from the Fock state vacuum $\ket{0}$ as follows 
\begin{equation}
\label{bog_vac}
\ket{\Omega}=\prod_q\gamma_{q+}\gamma_{q-}\ket{0}=\prod_q(u_q+v_qa^{\dagger}_{-q}a_q^{\dagger})\ket{0},
\end{equation}
which shows that the Bogoliubov vacuum contains the usual Fock vacuum, where a number of exotic  spinless Cooper pairs have been condensed into. One can check that in the limit of vanishing gap ($\Delta\to0$), the standard Fock vacuum is recovered. The positive- and negative-energy single-particle levels can be built from the Bogoliubov vacuum as $\ket{\epsilon_{q\pm}}=\gamma^{\dagger}_{q\pm}\ket{\Omega}$.



\begin{references}
\bibitem{comment1}
  In order to have adiabatically connected edge states, the number of sites in the original chain must be odd.
\end{references}
\end{document}